\documentclass{raa}           

\usepackage{graphicx,times}
\usepackage{natbib}
\usepackage{amssymb,amsmath}
\usepackage{longtable}
\usepackage{siunitx}
\bibpunct{(}{)}{;}{a}{}{,}

\usepackage{fontspec}
\usepackage{xeCJK}

\usepackage[pagebackref=true]{hyperref}

\begin{document}

  \title{Variation of the 2175 Å extinction feature in Andromeda galaxy}

 \volnopage{ {\bf 20XX} Vol.\ {\bf X} No. {\bf XX}, 000--000}
   \setcounter{page}{1}

   \author{Bing Yan (闫冰)\inst{1}, Shu Wang (王舒)\inst{3}\thanks{Corresponding author: shuwang@nao.cas.cn}, Jian Gao (高健)\inst{2,4}\thanks{Co-corresponding author: jiangao@bnu.edu.cn}, Yuxi Wang (王钰溪)\inst{5}, Bingqiu Chen (陈丙秋)\inst{1}\thanks{Co-corresponding author: bchen@ynu.edu.cn}
   }

   \institute{South-Western Institute for Astronomy Research, Yunnan University, Chenggong District, Kunming 650091, China; {\it bchen@ynu.edu.cn}\\
        \and
             Institute for Frontiers in Astronomy and Astrophysics, Beijing Normal University, Beijing 102206, China; {\it jiangao@bnu.edu.cn}\\
        \and
             CAS Key Laboratory of Optical Astronomy, National Astronomical Observatories, Chinese Academy of Sciences, Beijing 100101, Peopleʼs Republic of China; {\it shuwang@nao.cas.cn}\\
    	\and
             School of Physics and Astronomy, Beijing Normal University, Beijing 100875, China \\
        \and 
             Department of Astronomy, College of Physics and Electronic Engineering, Qilu Normal University, Jinan 250200, China; \\
\vs \no
   {\small Received 20XX Month Day; accepted 20XX Month Day}
}

\abstract{Extinction curves contain key information on interstellar dust composition and size distribution, with the 2175 Å bump being the most prominent feature. We analyze 20 sightlines toward M31 using HST/STIS UV spectroscopy combined with multi-band photometry to characterize this feature. The extinction curves show substantial diversity, from MW-like shapes to flatter profiles with $R_V$ reaching up to $\sim5.8$. The strength of the 2175 Å feature varies widely, including two sightlines where the bump is essentially absent. The bump central wavelength spans a broader range than previously reported, while its width remains consistent with earlier studies. A moderate positive correlation is found between bump strength ($c_3$) and width ($\gamma$). We derive an average UV extinction curve toward M31 with $R_V \approx 3.53$. These results provide new constraints on dust properties and their spatial variations in this galaxy. 
\keywords{ISM: dust, extinction --- stars: early-type --- ultraviolet: ISM --- ultraviolet: stars
}
}

   \authorrunning{B. Yan et al. }            
   \titlerunning{Variation of the 2175 Å extinction feature in M31}  
   \maketitle

%
\section{Introduction}           
\label{sect:intro}

Interstellar dust is a crucial component of the interstellar medium (ISM), as it absorbs and scatters stellar light, resulting in interstellar extinction. The wavelength dependence of extinction, expressed as the extinction curve, exhibits several characteristic features, including the ultraviolet (UV) bump at 2175 Å and silicate absorption features at 9.7 and 18 $\mu\text{m}$. The 2175 Å bump is the most prominent feature, with a nearly constant central wavelength but a variable width that depends on environment \citep{Fitzpatrick1986, Fitzpatrick1988, Fitzpatrick1990, Fitzpatrick2007}. 
Recent studies have suggested that polycyclic aromatic hydrocarbons (PAHs) may be the primary carriers of the 2175 Å feature \citep{Blasberger2017, Massa2022, Gordon2024, LinQi2023}, though definitive evidence is still lacking. 

The properties of interstellar extinction curves vary both among and within galaxies, with the 2175 Å bump being particularly sensitive to environment. In the Milky Way (MW), the bump is prominent, whereas the Magellanic Clouds (MCs) display more diverse behavior. The Large Magellanic Cloud (LMC) generally exhibits stronger far-UV extinction and a weaker bump, although the 30 Doradus region retains a relatively strong feature \citep{Fitzpatrick1985, Clayton1985}. In contrast, the Small Magellanic Cloud (SMC), particularly its Bar region, shows the steepest far-UV extinction and often lacks the 2175 Å feature entirely, though some regions display MW-like extinction with detectable bumps \citep{Gordon1998, Apellniz2012, Hagen2016, Gordon2024}.

Compared to the MW, the extinction curve of Andromeda galaxy (M31) exhibits a weaker 2175 Å bump but a similar overall slope \citep{Bianchi1996}. \citet{Dong2014} found that the extinction in M31 bulge ($R_V \approx 2.4-2.5$) resembles that of the Galactic bulge ($R_V \sim 2.5$) \citep{David2013}. More recently, \citet{Wang2022} demonstrated that M31’s extinction curves span a wide range of $R_V$ values, while the average curve is broadly similar to that of the MW, exhibiting a flatter far-UV rise, suggesting an ISM environment analogous to the diffuse regions of the MW. Previous studies of M31’s extinction curves have primarily relied on photometric data, which constrain the overall shape but provide limited insight into the properties and variations of the 2175 Å feature. Due to the inherent limitations of the photometric observation, spectroscopic observations are essential for a detailed characterization of this feature. To date, only \citet{Clayton2015} and \citet{Clayton2025} have analyzed 17 sightlines in M31 using UV spectra from the Hubble Space Telescope (HST). Their results show that the average extinction curve of M31 differs from those of the MW, LMC, and SMC, but closely resembles that of the LMC 30 Doradus region. 

\begin{figure}[htb!]
\centering
\includegraphics[width=0.8\textwidth]{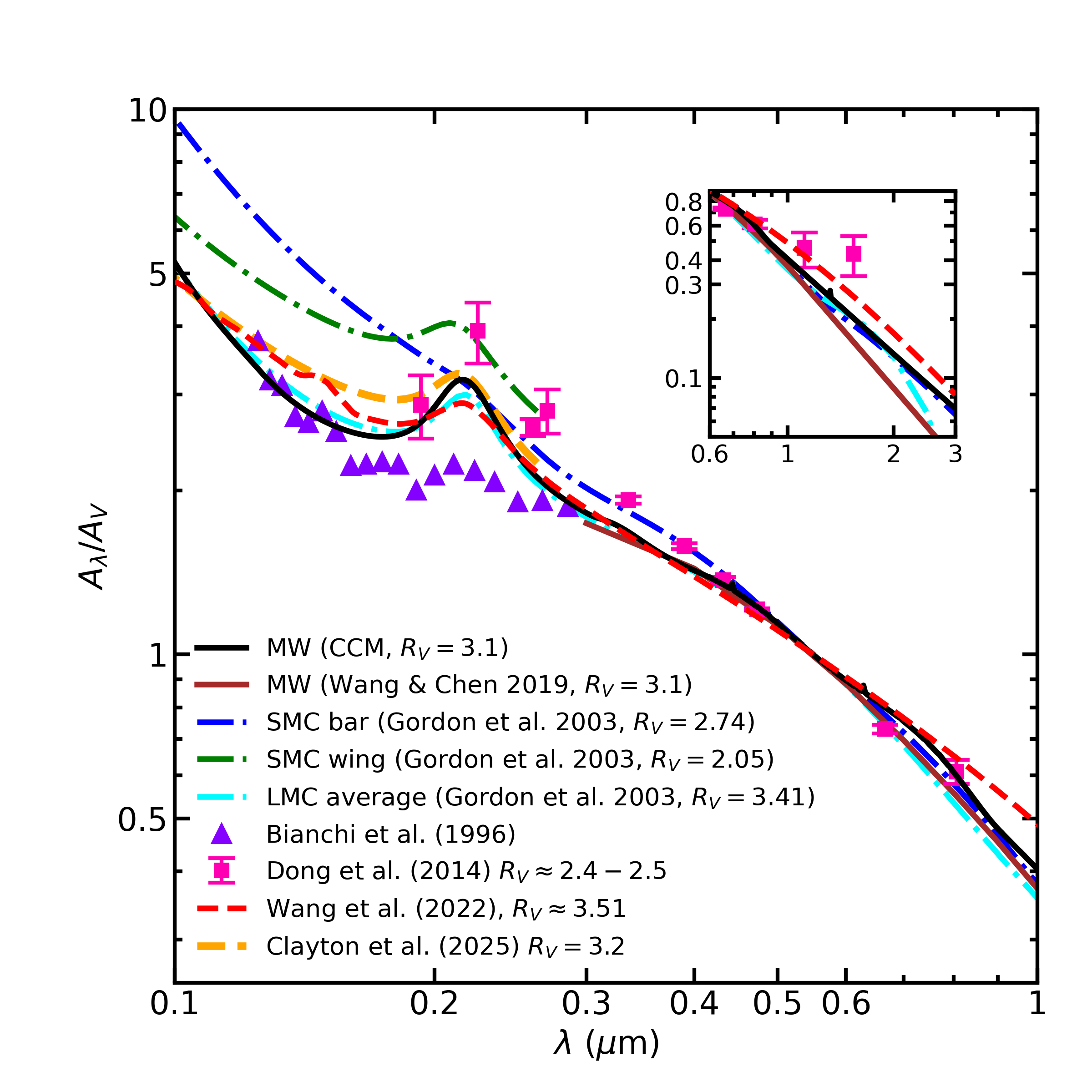}
\caption{Comparison of extinction curves for M31, the MW, and different regions of the MCs. The black and brown solid line denotes the average diffuse ISM curve of the MW from \citet{CCM89} and \citet{Wang2019}. The blue and green dotted lines correspond to the SMC Bar and Wing, respectively, while the cyan dotted line shows the LMC average from \citet{Gordon2003}. Violet triangles indicate the M31 extinction from \citet{Bianchi1996}, and magenta squares represent the average curves of the M31 bulge from \citet{Dong2014}. Orange and red dashed lines show the average curves from \citet{Clayton2025} and \citet{Wang2022}, respectively.}
\label{fig:extinction}
\end{figure}

Figure~\ref{fig:extinction} compares the extinction curves of M31, the MW, and different regions within the MCs. Except for the SMC Bar (blue dashed line), where the 2175 Å feature is absent, the average extinction curves in other regions display a pronounced bump. In M31, the curve toward the bulge (pink squares) is steeper and exhibits a stronger 2175 Å feature \citep{Dong2014}, whereas other sightlines show weaker bumps \citep{Bianchi1996, Clayton2025, Wang2022}. 

In this work, we focus on characterizing the 2175 Å feature along sightlines toward M31. Using extinction tracers from \citet{Wang2022} and \citet{Clayton2015}, we select 20 sightlines with HST UV spectroscopic data, including seven new targets, to examine variations in the bump strength and shape. This study offers one of the most detailed spectroscopic analyses of the 2175 Å feature in M31 to date, providing improved insights into interstellar dust properties and their spatial variations. The structure of the paper is as follows: Section 2 outlines the observational data, Section 3 describes the methodology, Section 4 presents and discusses the results, and Section 5 summarizes the main findings.

\section{Data and Sample}
\label{sec:data}

We selected a total of 147 O- and B-type supergiants in M31 from the studies of \citet{Wang2022} and \citet{Clayton2015} as our initial sample of extinction tracers. Among these, 20 stars have low-resolution UV spectra obtained with HST Space Telescope Imaging Spectrograph (STIS). For each object, we constructed a spectral energy distribution (SED) covering the UV to near-infrared (NIR) regime using multi-band photometry from the Local Group Galaxies Survey (LGGS), Pan-STARRS1 (PS1), United Kingdom Infrared Telescope (UKIRT), and Panchromatic Hubble Andromeda Treasury (PHAT), combined with UV data from Swift Ultraviolet-Optical Telescope (UVOT) and XMM-Newton Serendipitous Ultraviolet Source Survey (SUSS), as well as HST spectroscopy where available.

\subsection{Photometric data}
\citet{Wang2022} analyzed 140 O- and B-type supergiants in M31, from the LGGS catalog, to investigate the galaxy’s extinction law. \citet{Clayton2015} presented HST/STIS G140L and G230L observations for 11 early-type supergiants in M31 (see their Table 2). Combining these two samples and accounting for four overlaps yields 147 unique stars. 

Photometric data for these targets were compiled from multiple surveys: UBVRI bands from LGGS; NIR bands ($JHK$) from UKIRT Wide Field Camera (WFCAM); optical bands ($g, r, i, z, y$) from PS1 survey; six PHAT bands (F275W, F336W, F475W, F814W, F110W, F160W) spanning near-UV to NIR; three UV bands (UVW2, UVM2, UVW1) from Swift/UVOT; and another set of three UV bands (UVW2, UVM2, UVW1) from XMM-SUSS. Data acquisition and selection criteria follow those described in \citet{Wang2022} (see their Section 2 and Table 1).

\subsection{Spectral data}\label{specdata}

To identify stars with UV spectra, we cross-matched the 147 targets against the HST archive at Mikulski Archive for Space Telescopes (\href{https://archive.stsci.edu/missions-and-data/hst}{MAST}) using a $1 \arcsec$ radius, obtaining HST/STIS spectra for 20 stars. Observations were made with the G140L grating ($\sim1120-1715$ Å) and the G230L grating ($\sim1590-3150$ Å). During preprocessing, spurious negative flux values were removed, and the fitting ranges were restricted to $1230-1700$ Å and $1700-3100$ Å, respectively. To ensure consistency with theoretical models, all spectra were convolved to a uniform resolving power of $R=100$ using the \texttt{coronagraph} toolkit.

\begin{figure}[htb!]
\centering
\includegraphics[width=0.8\textwidth]{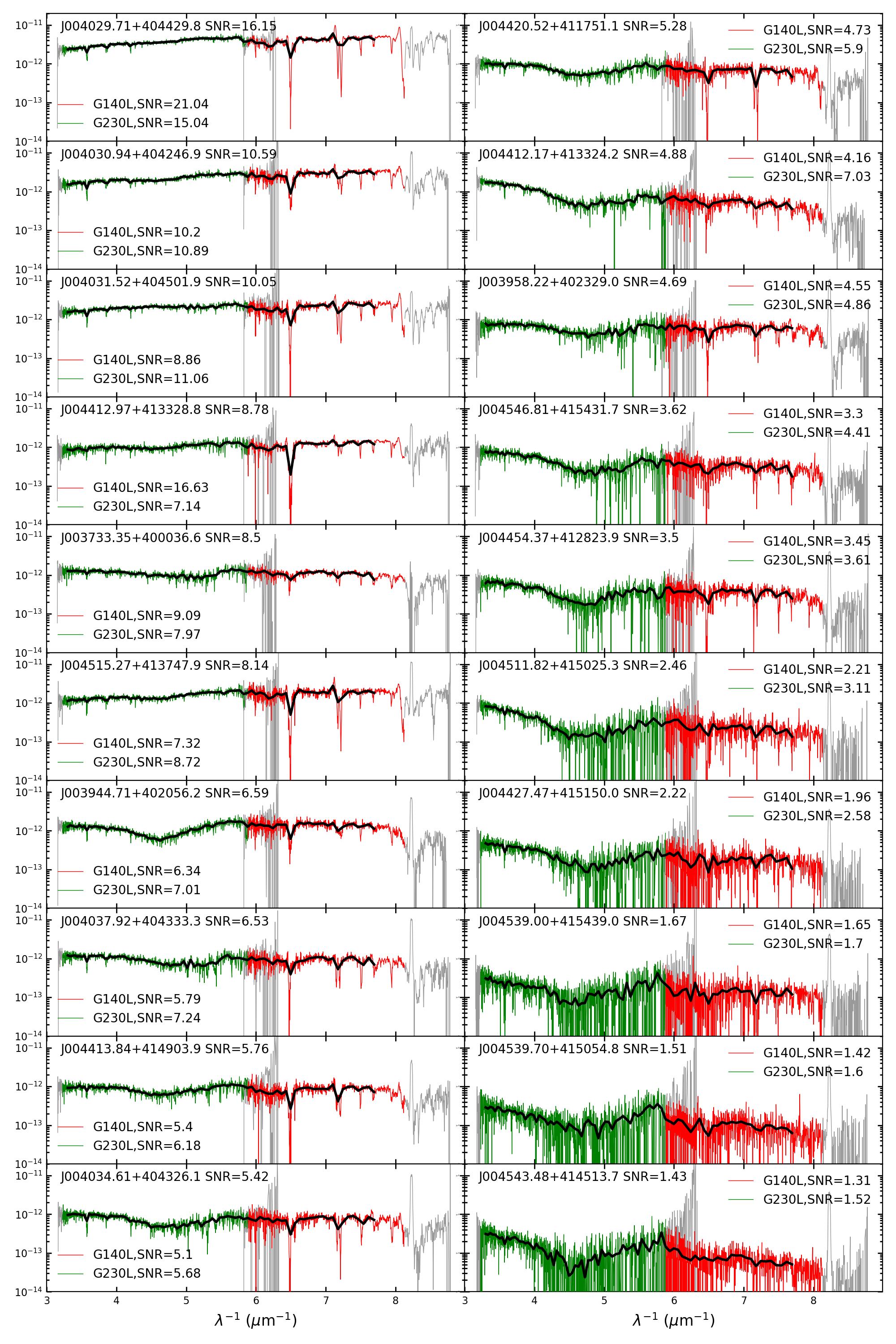}
\caption{HST/STIS UV spectra for the 20 sample stars. Red and green solid lines show the G140L and G230L grating spectra, respectively. Light-gray segments indicate regions excluded from the analysis, while black lines represent the spectra after resolution degradation to $R=100$ for fitting. The overall SNR of each spectrum is noted in the upper-left corner of each panel.}\label{fig:spec_all}
\end{figure}

Table~\ref{tab:spectrum} lists the spectroscopic information of the 20 stars, including the signal-to-noise ratio (SNR) for each grating. We further computed the total SNR (SNR$_\text{total}$) from the combined spectral segments as a measure of the overall spectral quality. Five stars with SNR$_\text{total}<3$ are identified as low-SNR sources and are marked with an asterisk in Table~\ref{tab:spectrum}. The fitting results for these stars are considered unreliable and are excluded from the subsequent analysis. Figure~\ref{fig:spec_all} displays the HST/STIS spectra: red and green solid lines indicate the original G140L and G230L data, respectively; gray segments denote trimmed portions; and black lines show the resolution-degraded spectra used for fitting.
{\footnotesize
\begin{longtable}{c c c c c c c S[table-format=2.2]}
\caption{HST/STIS Observations for the 20 sources in M31.}\label{tab:spectrum}\\
\hline\noalign{\smallskip}
Star & R.A.(J2000) & Decl.(J2000) & Date 
& Exp.(s) & Dataset & STIS & SNR \\
\hline\noalign{\smallskip}
\endfirsthead

\caption[]{HST/STIS Observations for the 20 sources in M31 (continued)}\\
\hline\noalign{\smallskip}
Star & R.A.(J2000) & Decl.(J2000) & Date 
& Exp.(s) & Dataset & STIS & SNR \\
\hline\noalign{\smallskip}
\endhead

\hline \multicolumn{7}{r}{{Continued on next page}} \\ \hline
\endfoot
\hline
\endlastfoot

J003733.35+400036.6\phantom{*} & 00 37 33.35 & +40 00 36.6 & 2000 Jul. 07 & 2323 & O56R12010 & G140L & 6.03\\
 & & & 2000 Jul. 07 & 3000 & O56R12020 & G140L & 6.80\\
 & & & 2012 Oct. 05 & 2423 & OBPX01010 & G230L & 7.97\\
J003944.71+402056.2\phantom{*} & 00 39 44.71 & +40 20 56.2 & 2013 Jul. 13 & 2494 & OBPX52010 & G140L & 6.34\\
 & & & 2013 Jan. 27 & 2908 & OBPX02020 & G140L & 4.66\\
 & & & 2013 Jan. 27 & 2423 & OBPX02010 & G230L & 7.00\\
J003958.22+402329.0\phantom{*} & 00 39 58.22 & +40 23 29.0 & 2013 Feb. 03 & 3077 & OBPX03020 & G140L & 4.55\\
 & & & 2013 Feb. 03 & 2423 & OBPX03010 & G230L & 4.86\\
J004029.71+404429.8\phantom{*} & 00 40 29.71 & +40 44 29.8 & 2004 Jan. 22 & 1820 & O8MG01010 & G140L & 9.51\\
 & & & 2004 Jan. 22 & 2840 & O8MG01020 & G140L & 11.98\\
 & & & 2004 Jan. 22 & 2800 & O8MG01030 & G140L & 8.30\\
 & & & 2004 Jan. 23 & 2800 & O8MG01040 & G140L & 8.15\\
 & & & 2004 Jan. 23 & 2800 & O8MG01050 & G140L & 8.25\\
 & & & 2012 Oct. 04 & 2423 & OBPX04010 & G230L & 15.04\\
J004030.94+404246.9\phantom{*} & 00 40 30.94 & +40 42 46.9 & 2013 Feb. 11 & 3077 & OBPX05020 & G140L & 10.20\\
 & & & 2013 Feb. 11 & 2423 & OBPX05010 & G230L & 10.89\\
J004031.52+404501.9\phantom{*} & 00 40 31.52 & +40 45 01.9 & 2012 Dec. 19 & 3077 & OBPX06020 & G140L & 8.86\\
 & & & 2012 Dec. 19 & 2423 & OBPX06010 & G230L & 11.06\\
J004034.61+404326.1\phantom{*} & 00 40 34.61 & +40 43 26.1 & 2013 Feb. 09 & 3077 & OBPX07020 & G140L & 5.10\\
 & & & 2013 Feb. 09 & 2423 & OBPX07010 & G230L & 5.68\\
J004037.92+404333.3\phantom{*} & 00 40 37.92 & +40 43 33.3 & 2013 Feb. 10 & 3077 & OBPX08020 & G140L & 5.79\\
 & & & 2013 Feb. 10 & 2423 & OBPX08010 & G230L & 7.24\\
J004412.17+413324.2\phantom{*} & 00 44 12.17 & +41 33 24.2 & 2012 Jun. 28 & 3077 & OBPX09020 & G140L & 4.15\\
 & & & 2012 Jun. 28 & 2423 & OBPX09010 & G230L & 7.03\\
J004412.97+413328.8\phantom{*} & 00 44 12.97 & +41 33 28.8 & 2003 Sep. 27 & 2200 & O8MG07010 & G140L & 5.89\\
 & & & 2003 Sep. 27 & 2840 & O8MG07020 & G140L & 6.66\\
 & & & 2003 Sep. 27 & 2800 & O8MG07030 & G140L & 4.60\\
 & & & 2003 Sep. 27 & 2800 & O8MG07040 & G140L & 4.57\\
 & & & 2003 Sep. 27 & 2800 & O8MG07050 & G140L & 4.41\\
 & & & 2003 Dec. 02 & 2200 & O8MG08010 & G140L & 5.88\\
 & & & 2003 Dec. 03 & 2840 & O8MG08020 & G140L & 6.47\\
 & & & 2003 Dec. 03 & 2800 & O8MG08030 & G140L & 4.42\\
 & & & 2003 Dec. 03 & 2800 & O8MG08040 & G140L & 4.41\\
 & & & 2003 Dec. 03 & 2800 & O8MG08050 & G140L & 4.46\\
 & & & 2012 Oct. 03 & 2423 & OBPX10010 & G230L & 7.14\\
J004413.84+414903.9\phantom{*} & 00 44 13.84 & +41 49 03.9 & 2017 Nov. 09 & 3043 & OD6J11020 & G140L & 5.40\\
 & & & 2017 Nov. 09 & 2357 & OD6J11010 & G230L & 6.19\\
J004420.52+411751.1\phantom{*} & 00 44 20.52 & +41 17 51.1 & 2017 Sep. 19 & 3043 & OD6J09020 & G140L & 4.73\\
 & & & 2017 Sep. 19 & 2357 & OD6J09010 & G230L & 5.90\\
J004427.47+415150.0* & 00 44 27.47 & +41 51 50.0 & 2017 Jul. 27 & 3043 & OD6J05020 & G140L & 1.96\\
 & & & 2017 Jul. 27 & 2357 & OD6J05010 & G230L & 2.58\\
J004454.37+412823.9\phantom{*} & 00 44 54.37 & +41 28 23.9 & 2017 Nov. 12 & 3043 & OD6J12020 & G140L & 3.45\\
 & & & 2017 Nov. 12 & 2357 & OD6J12010 & G230L & 3.61\\
J004511.82+415025.3* & 00 45 11.82 & +41 50 25.3 & 2017 Jul. 26 & 3043 & OD6J02020 & G140L & 2.21\\
 & & & 2017 Jul. 26 & 2357 & OD6J02010 & G230L & 3.11\\
J004515.27+413747.9\phantom{*} & 00 45 15.27 & +41 37 47.9 & 2013 Jan. 21 & 2494 & OBPX13010 & G140L & 7.32\\
 & & & 2012 Dec. 16 & 2423 & OBPX11010 & G230L & 8.72\\
J004539.00+415439.0* & 00 45 39.00 & +41 54 39.0 & 2017 Nov. 12 & 3043 & OD6J13020 & G140L & 1.65\\
 & & & 2017 Nov. 12 & 2357 & OD6J13010 & G230L & 1.70\\
J004539.70+415054.8* & 00 45 39.70 & +41 50 54.8 & 2017 Nov. 12 & 3043 & OD6J03020 & G140L & 1.42\\
 & & & 2017 Nov. 12 & 2357 & OD6J03010 & G230L & 1.60\\
J004543.48+414513.7* & 00 45 43.48 & +41 45 13.7 & 2017 Jul. 24 & 3043 & OD6J01020 & G140L & 1.31\\
 & & & 2017 Jul. 24 & 2357 & OD6J01010 & G230L & 1.52\\
J004546.81+415431.7\phantom{*} & 00 45 46.81 & +41 54 31.7 & 2017 Sep. 26 & 3043 & OD6J10020 & G140L & 3.30\\
 & & & 2017 Sep. 26 & 2357 & OD6J10010 & G230L & 4.41\\
\end{longtable}
}

\subsection{Data Selection}
To exclude foreground contamination, we applied parallax and proper-motion cuts based on Gaia Data Release 3 (DR3) following the criteria in \citet{Ren2021}. We also inspected PS1 images for crowding, confirming that no nearby sources lie within a $2\arcsec \times 2\arcsec$ region around any target. 
Because our primary goal is to investigate the 2175 Å extinction feature, all stars with HST spectra were retained. For stars without spectra, we required (i) at least three UV photometric bands ($\lambda<3000$ Å), and (ii) coverage on both sides of the 2175 Å feature. 

To facilitate analysis, we define an \textit{S2175} flag: \textit{S2175}=1 for stars with both spectroscopic and photometric data, and \textit{S2175}=0 for stars with photometry only. The final sample consists of 93 stars, including 20 \textit{S2175}=1 objects analyzed in detail in Section~\ref{sec:result}, and 73 \textit{S2175}=0 objects used for comparison.

\section{Method}
\label{sec:method}

\begin{figure}[htb!]
\centering
\includegraphics[width=0.7\textwidth]{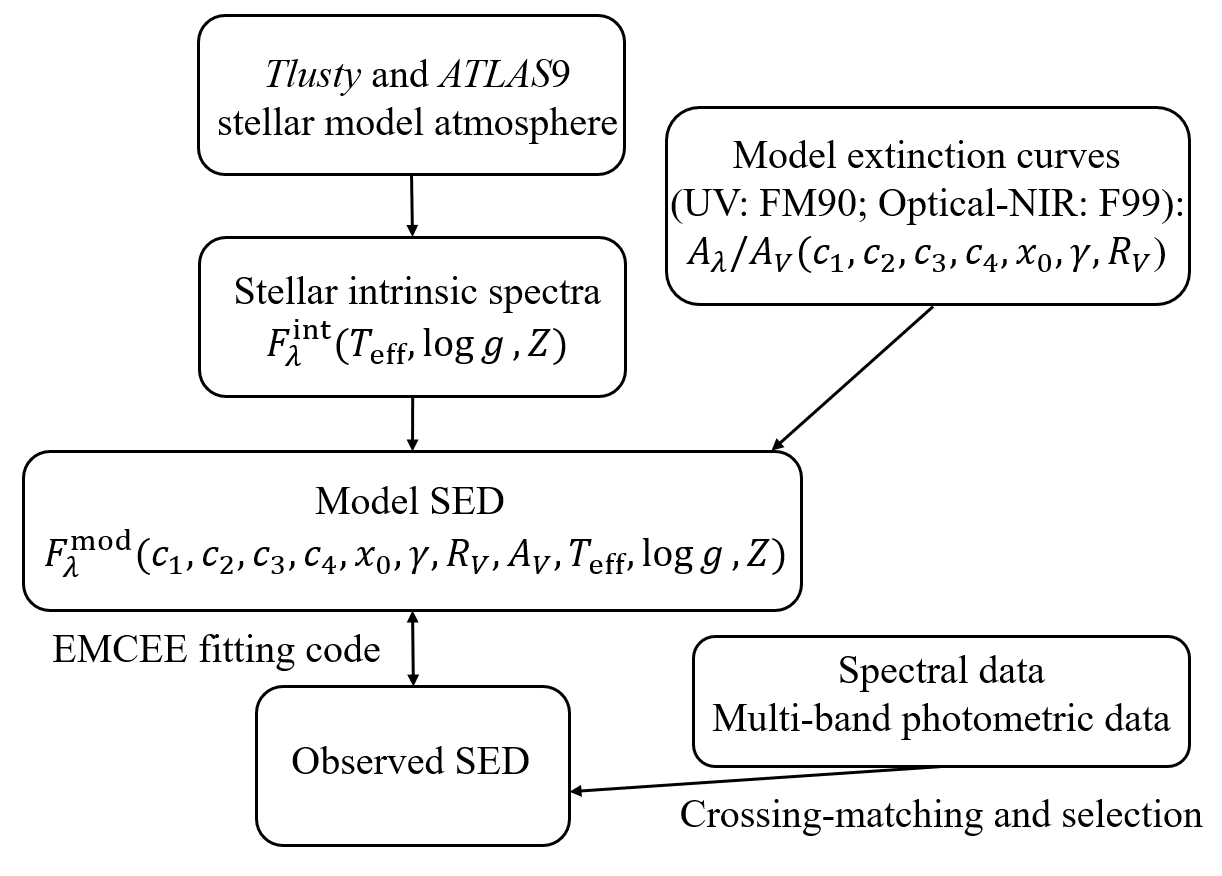}
\caption{The research workflow adopted in this study. Details of the observational data are provided in Section \ref{sec:data}. The extinction curve model and the intrinsic stellar spectra from atmospheric models are described in Sections \ref{sec:mod_ext} and \ref{sec:mod_intri}, respectively. The construction of the model SEDs and the fitting methodology are presented in Section \ref{sec:mod sed}.}\label{fig:method}
\end{figure}

Our methodology integrates the approaches of \citet{Clayton2015} and \citet{Wang2022}. We combine stellar atmosphere models with parameterized extinction laws to construct model SEDs and fit them to observed spectroscopic and photometric data using a Bayesian framework. The workflow is illustrated in Figure~\ref{fig:method}. Details of the observational data are given in Section \ref{sec:data}. Below, we describe the adopted extinction curves, the intrinsic stellar spectra, and the SED fitting procedure.

\subsection{The Extinction Curve} \label{sec:mod_ext}

We define $x \equiv \lambda^{-1}$, where $\lambda$ is in microns. For the UV regime ($x \ge 3.2~\mu\text{m}^{-1}$), the extinction curve is described using the FM90 parametrization \citep{Fitzpatrick1990}:
\begin{equation}
    E(x-V)/E(B-V) = c_1 + c_2x+c_3 D(x;\gamma,x_0)+c_4F(x),
\end{equation}
where the Drude profile is given by
\begin{equation}
    D(x;\gamma,x_0)=\frac{x^2}{(x^2-x_0^2)^2+x^2 \gamma^2},
\end{equation}
and the far-UV curvature term is
\begin{equation}
F(x)=
\begin{cases}
0.5392(x-5.9)^2 + 0.05644(x-5.9)^3, & x \ge 5.9~\mu\text{m}^{-1}, \\[3pt]
0, & x < 5.9~\mu\text{m}^{-1}.
\end{cases}
\end{equation}

The FM90 model consists of three main components: (1) a linear component defined by $c_1$ and $c_2$, setting the UV continuum slope and intercept; (2) a Drude profile characterized by $x_0$, $\gamma$, and $c_3$, representing the central wavelength, width, and strength of the 2175 Å bump; and (3) a far-UV curvature term parameterized by $c_4$. 
While $x_0$ is nearly constant among different sightlines, $\gamma$ exhibits significant variation. To reduce parameter degeneracy, we adopt the empirical relation $c_1 = 2.09 - 2.84c_2$ from \citet{Fitzpatrick2007}. The properties of the 2175 Å feature are quantified by integrated strength, $A_\text{bump} = \pi c_3 / (2\gamma R_V)$, and peak height above the continuum, $E_\text{bump} = c_3 / (\gamma^2 R_V)$.

For optical and NIR domains ($x < 3.2~\mu\text{m}^{-1}$), we use the F99 extinction law, with a single-parameter ($R_V$) \citep{Fitzpatrick1999}. This model is smoothly joined to the FM90 UV curve in the transition region ($2.9~\mu\text{m}^{-1}<x<3.5~\mu\text{m}^{-1}$) using a sigmoid function to ensure continuity. The F99 prescription is constructed by cubic spline interpolation through a set of anchor points that define the optical and NIR extinction behavior \citep{Fitzpatrick1999}.

\subsection{Intrinsic Stellar Spectral} \label{sec:mod_intri}

The intrinsic stellar spectra are derived using the nonlocal thermodynamic equilibrium (non-LTE) TLUSTY OSTAR and BSTAR grids, supplemented by the LTE ATLAS9 grid. In the overlapping parameter ranges, TLUSTY and ATLAS9 produce comparable spectra for the same stellar parameters; following \citet{Wang2022}, we adopt TLUSTY models for these regions to maintain consistency.

Based on the metallicity gradient of M31 reported by \citet{Liu2022}, the sample stars are expected to have metallicity in the range of approximately -0.2 to 0.3 dex, corresponding to $\sim 0.5-1.6~Z_\odot$. We estimated typical effective temperature ($T_\text{eff}$) and surface gravity ($\log g$) according to the spectral types of our sample stars. To assess the impact of metallicity on the UV spectra, we compared model spectra with 0.5, 1, and 2 $Z_\odot$, and found that the resulting differences across the UV range (1300–3000 Å) are minimal:
\begin{equation}
    \cfrac{\sqrt{\cfrac{1}{N}\textstyle \sum(F_\lambda(Z)-F_\lambda(Z=Z_\odot))^2}}{\cfrac{1}{N}\textstyle \sum F_\lambda(Z=Z_\odot)} \approx0.03
\end{equation}
This indicates that metallicity has a negligible effect on the UV continuum within our fitting range. Therefore, we adopt solar metallicity for all stars in the fitting process.

\subsection{SED Fitting} \label{sec:mod sed}

The observed stellar flux, $F_\lambda^\text{obs}$, is related to the intrinsic spectrum, $F_\lambda^\text{int}$, and extinction, $A_\lambda$, as \citep{Fitzpatrick2005}:
\begin{equation}
    F_\lambda^\text{obs}=F_\lambda^\text{int} \theta^2 10^{-0.4A_\lambda}，
\end{equation}
where $\theta=(R/d)^2$ represents the stellar angular diameter, with $d$ being the distance and $R$ the stellar radius.

For photometric data, the model spectra are convolved with the filter transmission curves to compute synthetic fluxes. For consistency with the observational data, the model spectra are degraded to a resolving power of $R = 100$, matching the resolution applied to the observed spectra as described in Section \ref{sec:data}. Combining the intrinsic spectra and extinction curve, the model flux is expressed as:
{\footnotesize
\begin{multline}\label{equ:Fmod}
    F_\lambda^{\text{mod}}(c_1,c_2,c_3,c_4,x_0,\gamma,R_V,A_V,T_{\text{eff}},\text{log}g,Z)
    =F_\lambda^{\text{int}}(T_{\text{eff}},\text{log}~g,Z)~\theta^2~10^{-0.4A_V\frac{A_\lambda}{A_V}(c_1,c_2,c_3,c_4,x_0,\gamma,R_V)}. 
\end{multline}
}

We employ the Markov Chain Monte Carlo (MCMC) technique using the \texttt{emcee} sampler \citep{Foreman2013} to explore the posterior distributions of model parameters. To improve efficiency while ensuring accuracy, a two-step fitting strategy is adopted. The first step performs a coarse grid search with large parameter intervals to identify approximate solutions. The second step refines the fit with smaller step sizes, fixing stellar parameters to the initial results.

\begin{table}[htb!]
\bc
\caption[]{Parameter Ranges and Steps for the Two-Step Fitting Process.}\label{tab:parameters}
\setlength{\tabcolsep}{3pt}
\begin{tabular}{c p{4.7cm} c c c c}
\hline\noalign{\smallskip}
 &  \centering First Step$^a$ & & & Second Step$^b$ &\\
Parameter & \centering Description & Range & Step Size &  Range & Step Size \\
\hline\noalign{\smallskip}
$T_\text{eff}$ & Effective Temperature of O-type supergiants (K) & [27500, 40000] & 2500 & - & - \\
 & Effective Temperature of B-type supergiants (K) & [10000, 30000] & 1000 & - & - \\
$\log g$ & Surface Gravity of O-type supergiants & [3.00, 3.50] & 0.25 & - & - \\
 & Surface Gravity of B-type supergiants & [2.25, 3.00] & 0.25 & - & - \\
$Z$ & Metallicity$^c$ & fixed, $Z_\odot$ & ... & - & ... \\
$A_V$ & V-band Extinction (mag) & [0.0, 4.0] & 1.0 & [$\overset{\frown}{A_V}-0.5, \overset{\frown}{A_V}+0.5$]$^d$ & 0.1 \\
$R_V$ & Total-to-selective extinction & [2.0, 6.0] & 1.0 & [$\overset{\frown}{R_V}-0.5, \overset{\frown}{R_V}+0.5$] & 0.1 \\
$c_2$ & UV Slope & [0.0, 1.5] & 0.5 & [$\overset{\frown}{c_2}-0.3, \overset{\frown}{c_2}+0.3$] & 0.1 \\
$c_3$ & 2175\,\AA\ Bump Height & [0.0, 6.0] & 1.0 & [$\overset{\frown}{c_3}-0.5, \overset{\frown}{c_3}+0.5$] & 0.1 \\
$c_4$ & far-UV Curvature & [-0.2, 2.0] & 0.44 & [$\overset{\frown}{c_4}-0.3, \overset{\frown}{c_4}+0.3$] & 0.1 \\
$x_0$ & 2175\,\AA\ Bump centroid & [4.0, 5.6] & 0.02 & [$\overset{\frown}{x_0}-0.03, \overset{\frown}{x_0}+0.03$] & 0.01 \\
$\gamma$ & 2175\,\AA\ Bump Width & [0.5, 1.5] & 0.2 & [$\overset{\frown}{\gamma}-0.3, \overset{\frown}{\gamma}+0.3$] & 0.1 \\
\hline
\end{tabular}
\ec
\tablecomments{0.9\textwidth}{
$^a$ Parameter ranges are based on the works of \citet{Clayton2015}, \citet{Wang2022}, and \citet{Wang2023}.\\
$^b$ The parameter with ``-'' is fixed to the result in the first fitting step.\\
$^c$ The metallicity is always fixed to one times the solar metallicity.\\
$^d$ $\overset{\frown}{X}$ denotes the first-step fitting result for parameter $X$.
}
\end{table}

The parameter ranges and step sizes are summarized in Table \ref{tab:parameters}, based on \citet{Clayton2015}, \citet{Wang2022}, and \citet{Wang2023}. Metallicity is fixed to the solar value for both steps, as discussed in Section \ref{sec:mod_intri}. We assume a Gaussian likelihood function and uniform priors for all nine free parameters. Convergence tests confirm consistency between the two steps, validating the robustness of this procedure.

Posterior distributions from \texttt{emcee} provide parameter estimates and uncertainties, with the median (50th percentile) adopted as the best-fit value and the 16th and 84th percentiles defining the $1\sigma$ confidence intervals.

\section{Result and Discussion}
\label{sec:result}
\subsection{The 2175Å Feature in M31}

\begin{figure}[htb!]
    \centering
    \includegraphics[width=0.7\textwidth]{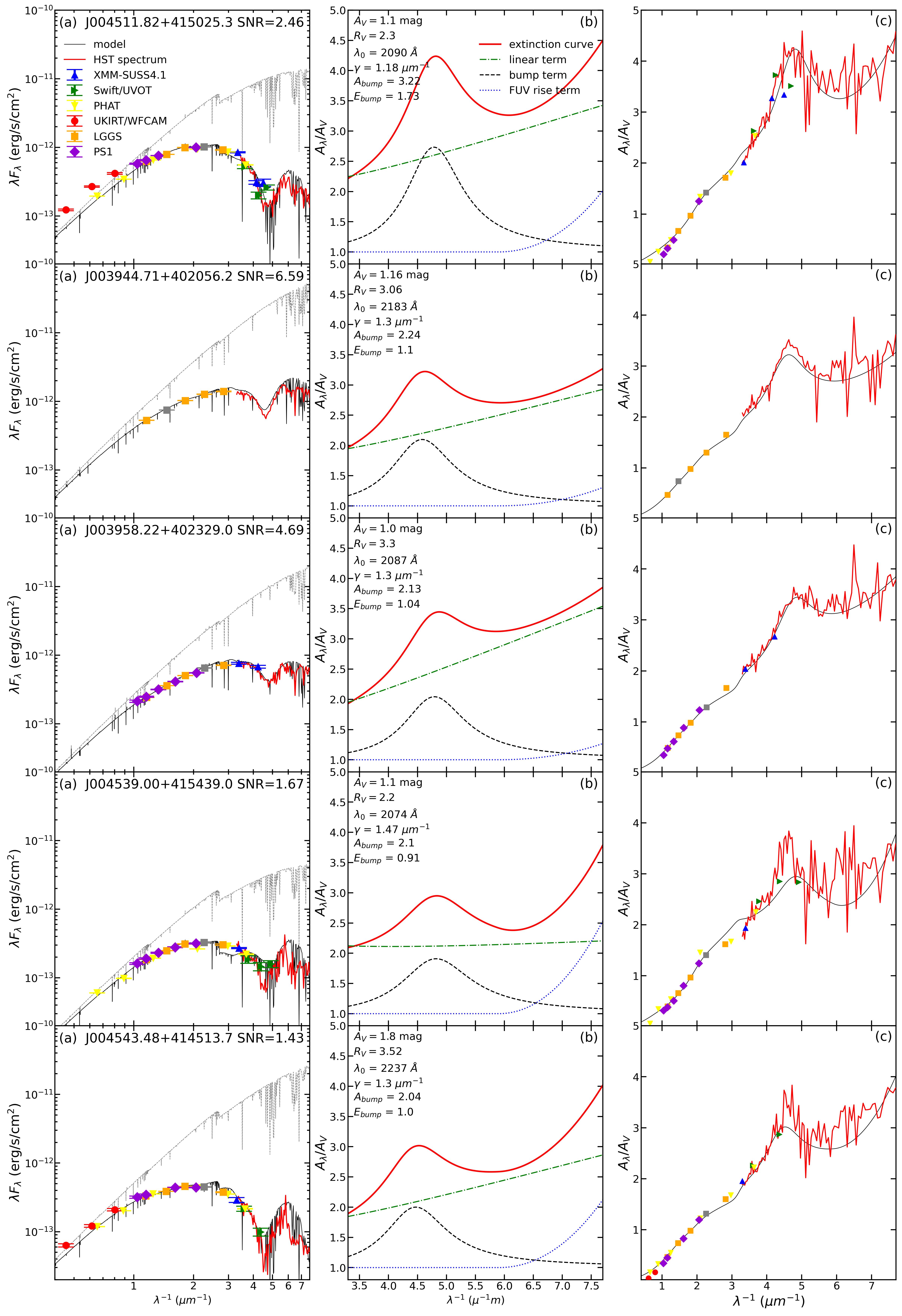}
    \caption{Observed spectra, intrinsic SEDs, and UV extinction curves for 20 sightlines with \textit{S2175}=1, ordered by decreasing $A_\text{bump}$. (a) Each panel lists the stellar ID and spectral S/N in the upper-left corner. The black solid line denotes the best-fit model spectrum, while the gray dashed line shows the intrinsic (unreddened) spectrum. The red solid line represents the observed HST spectrum, and the colored points correspond to photometric measurements in different bands. Gray points indicate data with missing uncertainties or those deviating by more than $3\sigma$ from the initial fit. (b) UV extinction curves, decomposed into individual components: total extinction (red solid line), linear background (green dotted line), 2175 Å bump (black dashed line), and far-UV rise (blue dotted line). (c) UV extinction curves compared with observed data, using the same plotting styles as in (a).}\label{fig:spec_NO0}
\end{figure}

\begin{figure}
    \centering
    \includegraphics[width=0.8\textwidth]{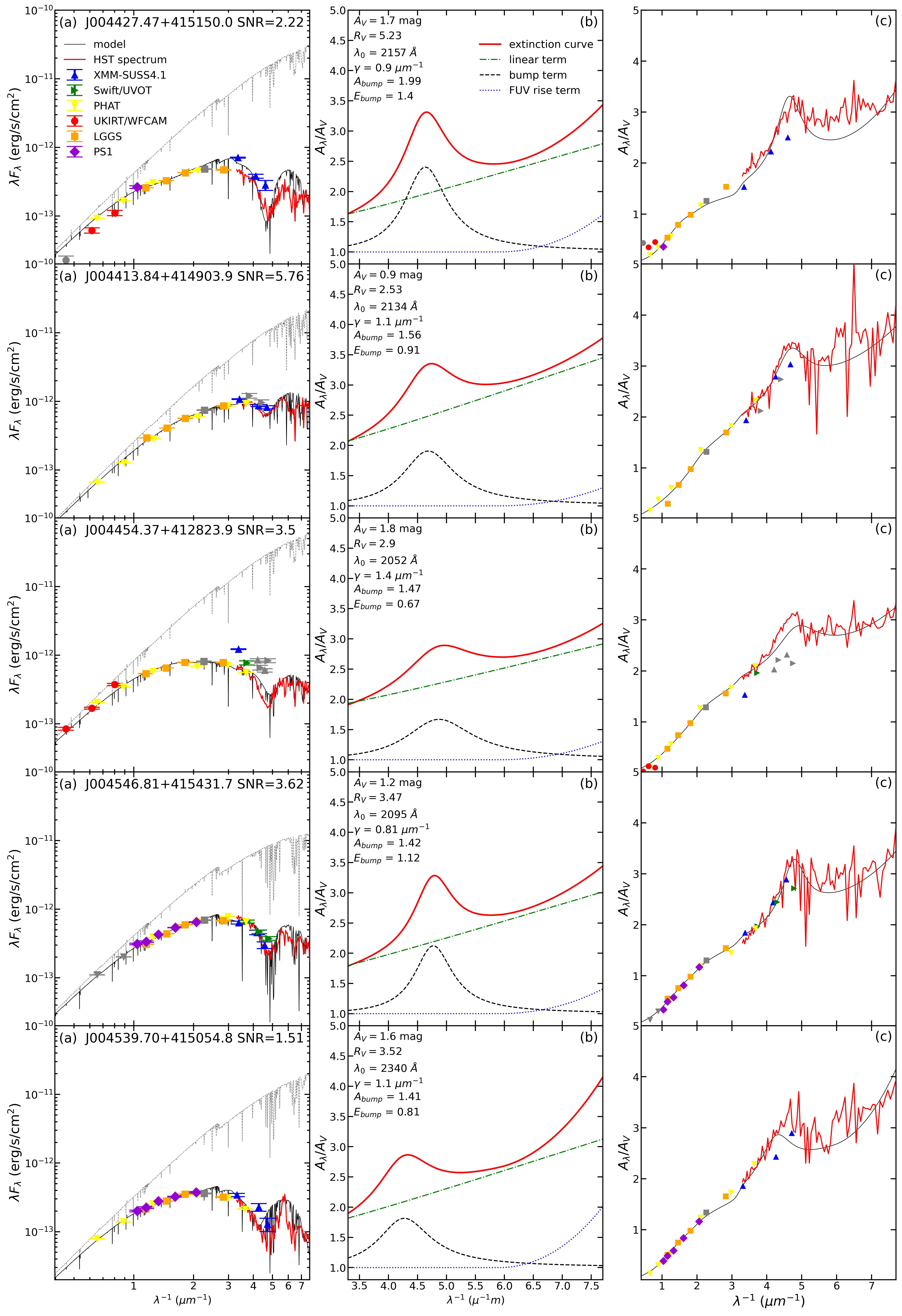}
    \addtocounter{figure}{-1}
    \caption{-- continued}
\end{figure}

\begin{figure}
    \centering
    \includegraphics[width=0.8\textwidth]{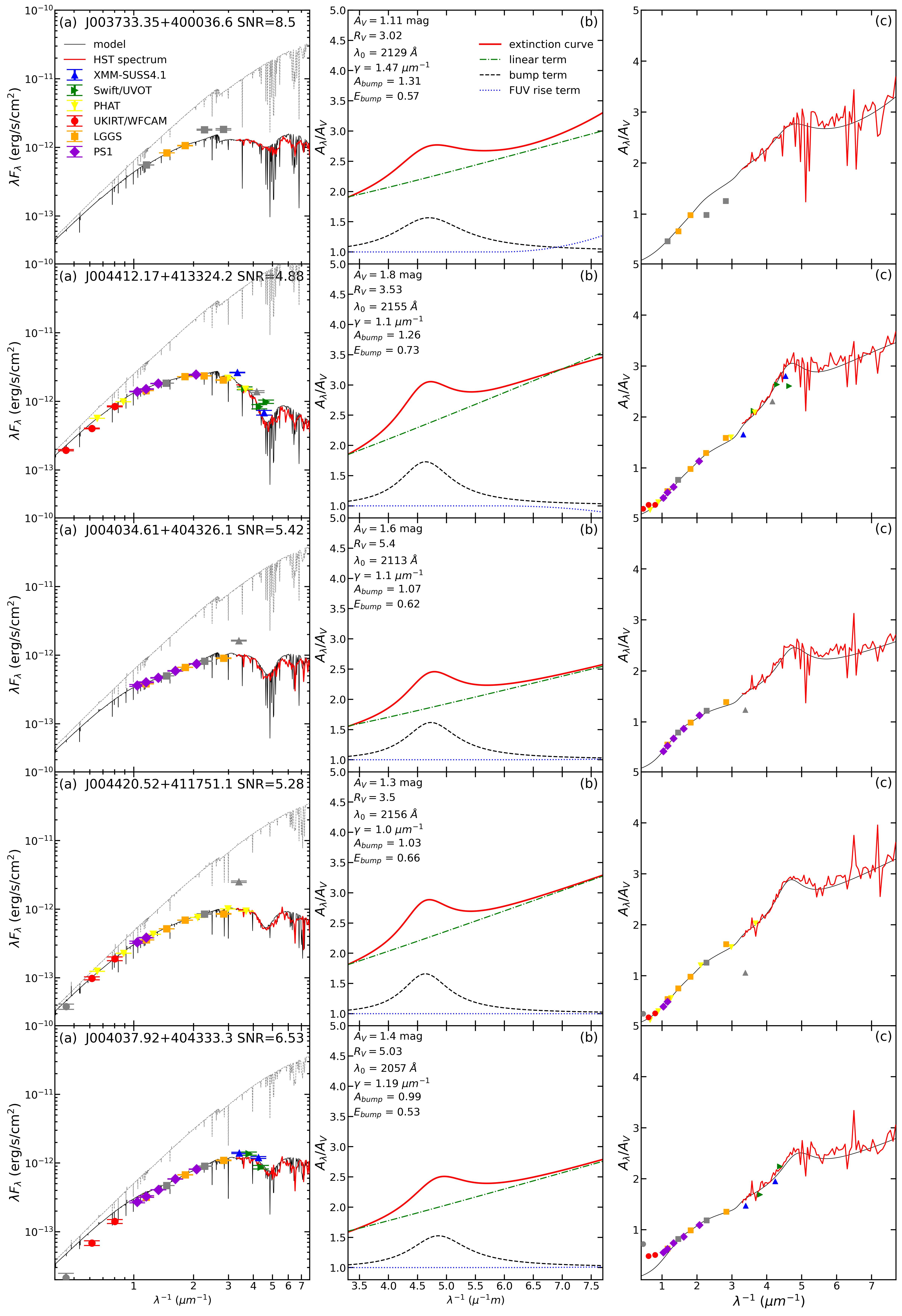}
    \addtocounter{figure}{-1}
    \caption{-- continued}
\end{figure}

\begin{figure}
    \centering
    \includegraphics[width=0.8\textwidth]{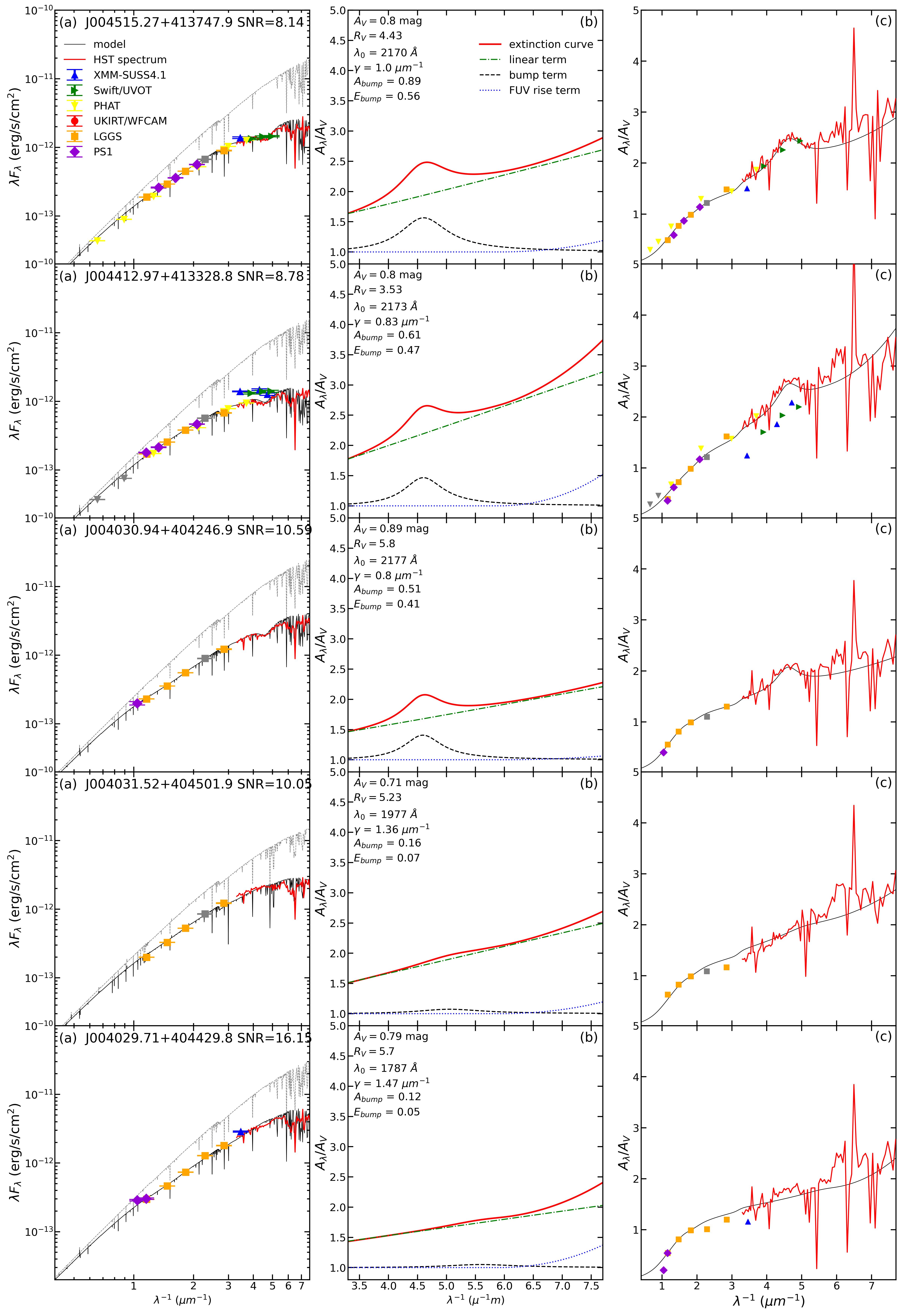}
    \addtocounter{figure}{-1}
    \caption{-- continued}
\end{figure}

\begin{table}[htb!]
\centering
\caption{List of stellar targets in M31 and derived parameters, including spectral types$~^a$}\label{tab:fit parameter}
\begin{tabular}{c c c c c c}
\hline\noalign{\smallskip}
LGGS ID & SpT. & $\log T_\text{eff}$ & $\log g$ & $A_V$ & $R_V$ \\
 & & K & dex & mag & \\
\hline\noalign{\smallskip}
J004511.82+415025.3 & B2Ia & $4.239_{-0.006}^{+0.057}$ & $2.31_{-0.04}^{+0.05}$ & $1.10_{-0.03}^{+0.03}$ & $2.30_{-0.03}^{+0.03}$\\
J003944.71+402056.2 & O9.7 Ib & $4.503_{-0.006}^{+0.007}$ & $3.32_{-0.05}^{+0.04}$ & $1.16_{-0.08}^{+0.06}$ & $3.06_{-0.37}^{+0.06}$\\
J003958.22+402329.0 & B0.7Ia & $4.473_{-0.003}^{+0.003}$ & $2.95_{-0.04}^{+0.04}$ & $1.00_{-0.03}^{+0.03}$ & $3.30_{-0.03}^{+0.03}$\\
J004539.00+415439.0 & B2.5I: & $4.229_{-0.069}^{+0.012}$ & $2.57_{-0.05}^{+0.04}$ & $1.10_{-0.03}^{+0.03}$ & $2.20_{-0.03}^{+0.03}$\\
J004543.48+414513.7 & B2I & $4.337_{-0.004}^{+0.003}$ & $2.94_{-0.04}^{+0.04}$ & $1.80_{-0.03}^{+0.03}$ & $3.52_{-0.02}^{+0.02}$\\
J004427.47+415150.0 & B1.5Ia & $4.427_{-0.003}^{+0.003}$ & $2.94_{-0.05}^{+0.04}$ & $1.70_{-0.03}^{+0.03}$ & $5.23_{-0.02}^{+0.02}$\\
J004413.84+414903.9 & OC9.7Ia & $4.486_{-0.006}^{+0.006}$ & $3.32_{-0.05}^{+0.04}$ & $0.90_{-0.03}^{+0.03}$ & $2.53_{-0.02}^{+0.02}$\\
J004454.37+412823.9 & B1Ie & $4.427_{-0.003}^{+0.003}$ & $2.94_{-0.05}^{+0.04}$ & $1.80_{-0.04}^{+0.04}$ & $2.90_{-0.04}^{+0.04}$\\
J004546.81+415431.7 & B2.5I & $4.285_{-0.004}^{+0.004}$ & $2.56_{-0.04}^{+0.05}$ & $1.20_{-0.04}^{+0.04}$ & $3.47_{-0.02}^{+0.02}$\\
J004539.70+415054.8 & B2.5I & $4.376_{-0.003}^{+0.003}$ & $2.93_{-0.05}^{+0.04}$ & $1.60_{-0.03}^{+0.03}$ & $3.52_{-0.02}^{+0.02}$\\
J003733.35+400036.6 & B2 Ia & $4.333_{-0.072}^{+0.007}$ & $2.88_{-0.33}^{+0.08}$ & $1.11_{-0.04}^{+0.07}$ & $3.02_{-0.04}^{+0.25}$\\
J004412.17+413324.2 & B3Ia+Neb & $4.376_{-0.003}^{+0.003}$ & $2.94_{-0.04}^{+0.04}$ & $1.80_{-0.03}^{+0.03}$ & $3.53_{-0.02}^{+0.02}$\\
J004034.61+404326.1 & B1 Ia & $4.375_{-0.004}^{+0.003}$ & $2.93_{-0.05}^{+0.05}$ & $1.60_{-0.04}^{+0.03}$ & $5.40_{-0.04}^{+0.04}$\\
J004420.52+411751.1 & B0.5Ia & $4.474_{-0.003}^{+0.044}$ & $2.95_{-0.04}^{+0.04}$ & $1.30_{-0.04}^{+0.04}$ & $3.50_{-0.04}^{+0.04}$\\
J004037.92+404333.3 & B1.5Ia & $4.425_{-0.031}^{+0.005}$ & $2.93_{-0.05}^{+0.05}$ & $1.40_{-0.03}^{+0.03}$ & $5.03_{-0.02}^{+0.02}$\\
J004515.27+413747.9 & O8 I & $4.538_{-0.006}^{+0.006}$ & $3.44_{-0.05}^{+0.05}$ & $0.80_{-0.03}^{+0.03}$ & $4.43_{-0.02}^{+0.02}$\\
J004412.97+413328.8 & O8.5Ia(f) & $4.520_{-0.006}^{+0.006}$ & $3.43_{-0.04}^{+0.05}$ & $0.80_{-0.03}^{+0.03}$ & $3.53_{-0.02}^{+0.02}$\\
J004030.94+404246.9 & O9.5 Ib & $4.504_{-0.006}^{+0.006}$ & $3.43_{-0.05}^{+0.05}$ & $0.89_{-0.04}^{+0.04}$ & $5.80_{-0.04}^{+0.04}$\\
J004031.52+404501.9 & B0.5 Ia & $4.427_{-0.003}^{+0.003}$ & $2.93_{-0.05}^{+0.05}$ & $0.71_{-0.04}^{+0.05}$ & $5.23_{-0.02}^{+0.38}$\\
J004029.71+404429.8 & O7-7.5 Iaf & $4.520_{-0.006}^{+0.006}$ & $3.43_{-0.05}^{+0.05}$ & $0.79_{-0.04}^{+0.04}$ & $5.70_{-0.16}^{+0.05}$\\
\hline\noalign{\smallskip}
\end{tabular}

\vspace{0.5em}

\begin{tabular}{c c c c c c c c c}
  \hline\noalign{\smallskip}
$c_2$ & $c_3$ & $c_4$ & $x_0$ & $\gamma$ & E(B-V) & $A_\text{bump}$ & $E_\text{bump}$ & \textit{S2175} \\
 & & & $\mu \text{m}^{-1}$ & $\mu \text{m}^{-1}$ & mag & & & \\
  \hline\noalign{\smallskip}
$0.70_{-0.03}^{+0.03}$ & $5.59_{-0.36}^{+0.06}$ & {\phantom{$-$}$1.13_{-0.02}^{+0.02}$} & $4.785_{-0.003}^{+0.007}$ & $1.18_{-0.06}^{+0.04}$ & 0.478  & 3.224  & 1.733 & 1\\
$0.76_{-0.04}^{+0.06}$ & $5.66_{-0.18}^{+0.07}$ & {\phantom{$-$}$0.45_{-0.03}^{+0.07}$} & $4.581_{-0.012}^{+0.007}$ & $1.30_{-0.03}^{+0.03}$ & 0.380  & 2.239  & 1.098 & 1\\
$1.27_{-0.02}^{+0.02}$ & $5.81_{-0.04}^{+0.06}$ & {\phantom{$-$}$0.43_{-0.02}^{+0.02}$} & $4.793_{-0.002}^{+0.002}$ & $1.30_{-0.03}^{+0.03}$ & 0.303  & 2.129  & 1.042 & 1\\
$0.10_{-0.03}^{+0.03}$ & $4.34_{-0.03}^{+0.07}$ & {\phantom{$-$}$1.63_{-0.02}^{+0.02}$} & $4.822_{-0.008}^{+0.012}$ & $1.47_{-0.02}^{+0.02}$ & 0.500  & 2.103  & 0.908 & 1\\
$0.90_{-0.03}^{+0.03}$ & $5.96_{-0.08}^{+0.03}$ & {\phantom{$-$}$1.91_{-0.13}^{+0.07}$} & $4.471_{-0.009}^{+0.009}$ & $1.30_{-0.03}^{+0.03}$ & 0.510  & 2.042  & 1.000 & 1\\
$1.48_{-0.02}^{+0.02}$ & $5.97_{-0.02}^{+0.02}$ & {\phantom{$-$}$1.56_{-0.07}^{+0.06}$} & $4.636_{-0.004}^{+0.006}$ & $0.90_{-0.04}^{+0.04}$ & 0.325  & 1.989  & 1.403 & 1\\
$0.83_{-0.06}^{+0.08}$ & $2.77_{-0.25}^{+0.06}$ & {\phantom{$-$}$0.37_{-0.27}^{+0.06}$} & $4.686_{-0.010}^{+0.003}$ & $1.10_{-0.03}^{+0.04}$ & 0.356  & 1.565  & 0.906 & 1\\
$0.70_{-0.03}^{+0.03}$ & $3.80_{-0.04}^{+0.04}$ & {\phantom{$-$}$0.43_{-0.02}^{+0.02}$} & $4.872_{-0.002}^{+0.002}$ & $1.40_{-0.03}^{+0.03}$ & 0.621  & 1.471  & 0.669 & 1\\
$1.00_{-0.04}^{+0.04}$ & $2.53_{-0.02}^{+0.05}$ & {\phantom{$-$}$0.69_{-0.07}^{+0.04}$} & $4.774_{-0.007}^{+0.009}$ & $0.81_{-0.04}^{+0.04}$ & 0.345  & 1.420  & 1.122 & 1\\
$1.10_{-0.03}^{+0.03}$ & $3.47_{-0.02}^{+0.02}$ & {\phantom{$-$}$1.71_{-0.10}^{+0.06}$} & $4.273_{-0.002}^{+0.002}$ & $1.10_{-0.03}^{+0.03}$ & 0.454  & 1.408  & 0.815 & 1\\
$0.80_{-0.03}^{+0.04}$ & $3.70_{-0.05}^{+0.05}$ & {\phantom{$-$}$0.40_{-0.04}^{+0.04}$} & $4.697_{-0.002}^{+0.002}$ & $1.47_{-0.02}^{+0.02}$ & 0.370  & 1.309  & 0.566 & 1\\
$1.40_{-0.04}^{+0.03}$ & $3.10_{-0.04}^{+0.04}$ & $-0.17_{-0.02}^{+0.02}$ & $4.640_{-0.004}^{+0.004}$ & $1.10_{-0.03}^{+0.03}$ & 0.511  & 1.258  & 0.729 & 1\\
$1.27_{-0.02}^{+0.02}$ & $4.02_{-0.06}^{+0.08}$ & {\phantom{$-$}$0.03_{-0.02}^{+0.02}$} & $4.733_{-0.002}^{+0.002}$ & $1.10_{-0.03}^{+0.03}$ & 0.297  & 1.065  & 0.616 & 1\\
$1.20_{-0.03}^{+0.03}$ & $2.30_{-0.04}^{+0.04}$ & {\phantom{$-$}$0.00_{-0.04}^{+0.04}$} & $4.637_{-0.002}^{+0.002}$ & $1.00_{-0.03}^{+0.03}$ & 0.371  & 1.034  & 0.659 & 1\\
$1.37_{-0.02}^{+0.02}$ & $3.77_{-0.08}^{+0.02}$ & {\phantom{$-$}$0.03_{-0.02}^{+0.02}$} & $4.863_{-0.002}^{+0.002}$ & $1.19_{-0.04}^{+0.04}$ & 0.279  & 0.986  & 0.526 & 1\\
$1.10_{-0.03}^{+0.03}$ & $2.50_{-0.03}^{+0.03}$ & {\phantom{$-$}$0.40_{-0.03}^{+0.03}$} & $4.607_{-0.008}^{+0.006}$ & $1.00_{-0.03}^{+0.03}$ & 0.181  & 0.887  & 0.565 & 1\\
$1.17_{-0.02}^{+0.02}$ & $1.13_{-0.05}^{+0.20}$ & {\phantom{$-$}$0.89_{-0.33}^{+0.35}$} & $4.603_{-0.002}^{+0.002}$ & $0.83_{-0.05}^{+0.08}$ & 0.227  & 0.607  & 0.466 & 1\\
$1.01_{-0.04}^{+0.04}$ & $1.50_{-0.04}^{+0.04}$ & {\phantom{$-$}$0.17_{-0.02}^{+0.02}$} & $4.594_{-0.007}^{+0.008}$ & $0.80_{-0.04}^{+0.04}$ & 0.154  & 0.510  & 0.407 & 1\\
$1.17_{-0.02}^{+0.02}$ & $0.72_{-0.48}^{+0.09}$ & {\phantom{$-$}$0.49_{-0.13}^{+0.04}$} & $5.057_{-0.002}^{+0.002}$ & $1.36_{-0.15}^{+0.02}$ & 0.136  & 0.158  & 0.073 & 1\\
$0.77_{-0.02}^{+0.02}$ & $0.63_{-0.05}^{+0.32}$ & {\phantom{$-$}$1.02_{-0.05}^{+0.12}$} & $5.597_{-0.002}^{+0.002}$ & $1.47_{-0.02}^{+0.02}$ & 0.139  & 0.117  & 0.050 & 1\\
\hline\noalign{\smallskip}
\end{tabular}
\vspace{0.7em}

\tablecomments{0.86\textwidth}{$^a$ This is a table of excerpts from the results of the \textit{S2175}=1 sample. The entire table，which also includes the \textit{S2175}=0 sample, is available in machine-readable form.}
\end{table}

Figure~\ref{fig:spec_NO0} shows the fitting results for 20 sightlines classified with \textit{S2175} = 1. The corresponding best-fit parameters are listed in Table~\ref{tab:fit parameter}, where stellar IDs and spectral types (SpT.) are adopted from the LGGS catalog \citep{Massey2016}. For five stars marked with an asterisk in Table~\ref{tab:spectrum}, the derived bump parameters ($c_3$, $c_4$, $x_0$, $\gamma$) are considered unreliable due to spectral distortions caused by low SNR. The five stars are therefore excluded from the subsequent discussion.

\begin{figure}[htb!]
    \centering
    \includegraphics[width=0.7\textwidth]{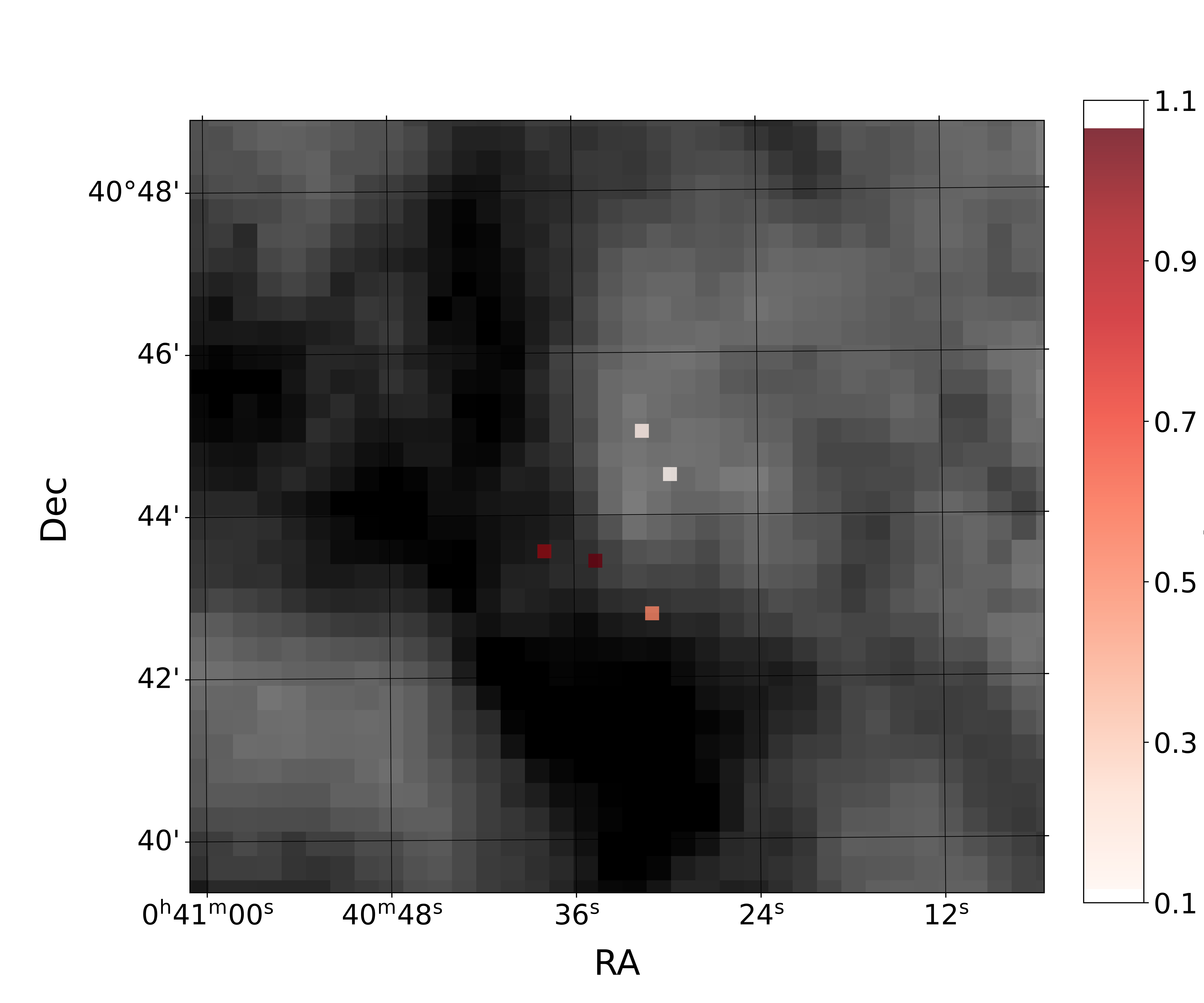}
    \caption{Distribution of the bump strength parameter $A_\text{bump}$ for five closely spaced sightlines, overlaid on the dust mass surface density map from \citet{Draine2014}. The three sightlines that exhibit a detectable 2175 Å feature are located along the edges of dense dust clump, whereas the two sightlines lacking the feature are located in an adjacent low-density areas.}
    \label{fig:fivestars}
\end{figure}

For the sightlines analyzed here, two sightlines exhibit an almost absent 2175 Å feature, with $A_\text{bump} < 0.2$, $E_\text{bump} < 0.1$, and $c_3 < 1$ (see the bottom two panels of Figure~\ref{fig:spec_NO0}). These sightlines correspond to stars J004031.52+404501.9 and J004029.71+404429.8, both of which have high-SNR HST spectra (SNR $>$ 10), lending strong confidence to the fits. Owing to the almost absent 2175 Å feature, these two stars are excluded from the discussion of its parameters in Section~\ref{sec:parameter}. J004031.52+404501.9 is a B-type supergiant with derived values of $\log T_\text{eff} = 4.43$ and $\log g = 2.93$, while J004029.71+404429.8 is an O-type supergiant with derived values of $\log T_\text{eff} = 4.52$ and $\log g = 3.43$. These two stars share similar extinction properties, including relatively small $A_V$ values (0.71 and 0.79 mag) and color excesses [$E(B-V) = 0.136$ and 0.139 mag], as well as high $R_V$ (5.23 and 5.70). Moreover, they are located within an angular separation of $\sim38 \arcsec$ and both lack a prominent 2175 Å feature, suggesting that they may reside in a common local environment within M31. In addition, three other sightlines(J004030.94+404246.9,J004034.61+404326.1,J004037.92+404333.3) in our sample are located in close proximity (within $\sim2\arcmin$) to the two bump-less sightlines. To investigate the possible environmental origin of the missing 2175 Å feature, we examined the local dust and gas conditions using the dust surface density map from \citet{Draine2014}, the CO emission map from \citet{Nieten2006}, and the H I distribution from \citet{Braun2009}. We find that the three sightlines exhibiting the 2175 Å feature lie along the edges of regions with relatively high dust, CO, and H I surface densities, while the two bump-less sightlines fall just outside these regions, in areas of noticeably lower dust and gas content. As shown in Figure~\ref{fig:fivestars}, these two featureless sightlines are located in a cavity adjacent to a dense dust clump. All five sightlines show higher-than-median SNRs and their extinction curve fits match the observed spectra well, reducing the likelihood that the absence of the 2175 Å feature is due to observational noise or fitting artifacts. Notably, one of the nearby sightlines, J004034.61+404326.1, also shows a weak bump and was independently analyzed by \citet{Clayton2025}(ID e22), supporting the consistency of our results. Therefore, we suggest that the absence or weakness of the 2175 Å feature in these sightlines is most likely caused by the locally reduced dust and gas densities, rather than limitations in the observations or fitting uncertainties.

Excluding the five low-SNR stars and the two stars with an almost absent feature, the central wavelength ($\lambda_0 = x_0^{-1}$) of the 2175 Å bump for the remaining 13 stars shows considerable variation, ranging from $\sim$ 2052 to 2183 Å. The bump width ($\gamma$) spans $0.8$–$1.47~\mu\text{m}^{-1}$. Despite the limited number of sightlines with spectroscopic data, the 2175 Å feature exhibits remarkable diversity, with bump strength parameters spanning $E_\text{bump} \sim 0.41–1.12$ and $A_\text{bump} \sim 0.51–2.23$.

\begin{figure}[htb!]
    \centering
    \includegraphics[width=0.6\textwidth]{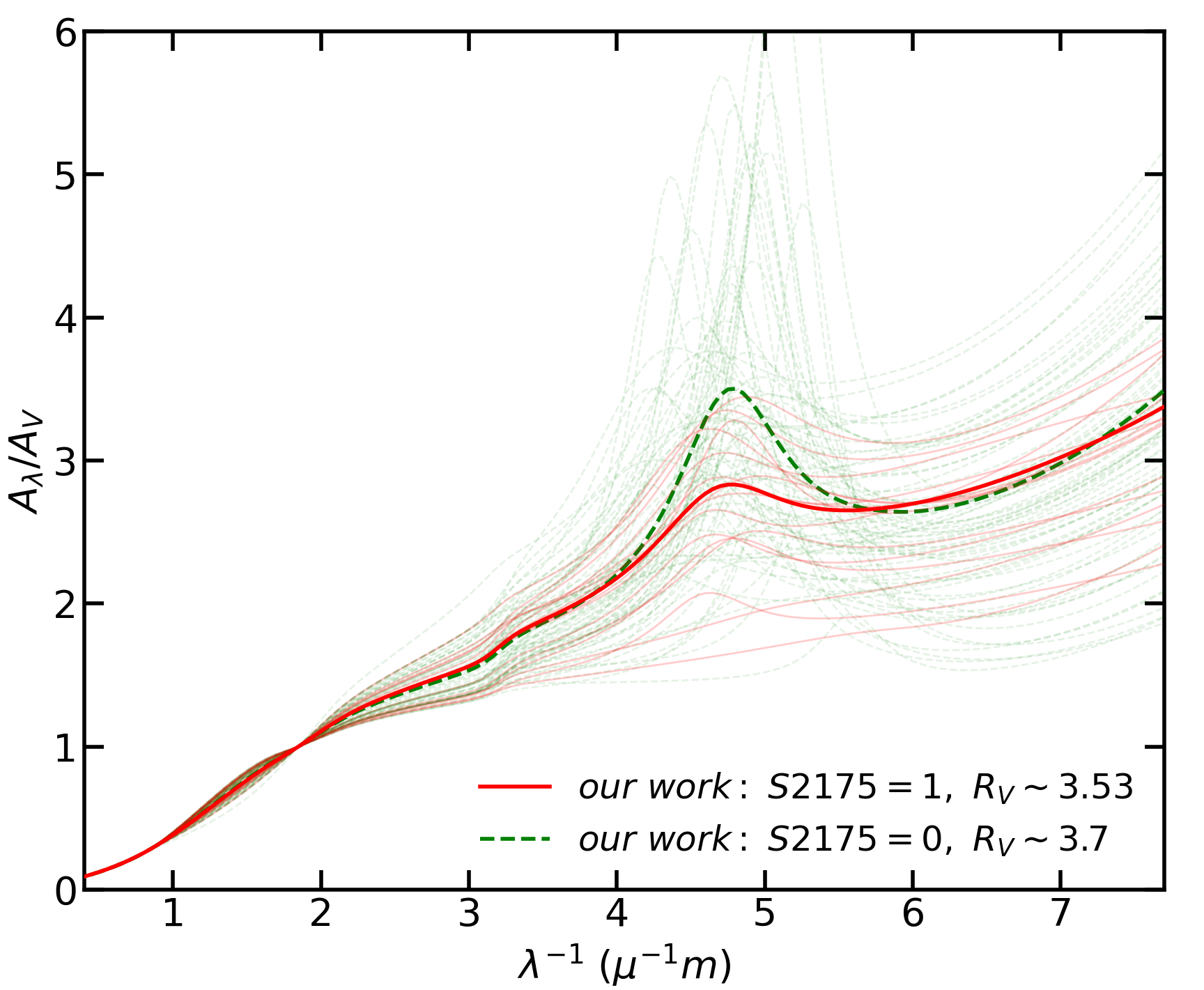}
    \caption{Comparison of extinction curves for 15 sources with \textit{S2175}=1 (red; individual curves in light red) and 75 sources with \textit{S2175}=0 (green; individual curves in light green). Photometry-only sources yield weaker constraints on the 2175 Å feature and occasionally produce artificially narrow, prominent bumps.}
    \label{fig:compare_spec&phm}
\end{figure}

To investigate possible systematic differences, we compare in Figure~\ref{fig:compare_spec&phm} the extinction curves of the \textit{S2175} = 1 group (excluding the five low-SNR stars) with those of the \textit{S2175} = 0 group. Both groups follow similar overall trends, with mean $R_V$ values of 3.53 and 3.7, respectively. The 2175 Å bump is weaker on average in the \textit{S2175} = 1 group, whereas some \textit{S2175} = 0 sightlines show anomalous shapes due to incomplete UV coverage. These anomalies disappear when at least seven UV photometric bands are included ($n_\text{UV band} = 7$), highlighting the necessity of either broad UV photometry or spectroscopy to robustly characterize the 2175 Å feature.

\begin{figure}[htb!]
    \centering
    \includegraphics[width=0.7\textwidth]{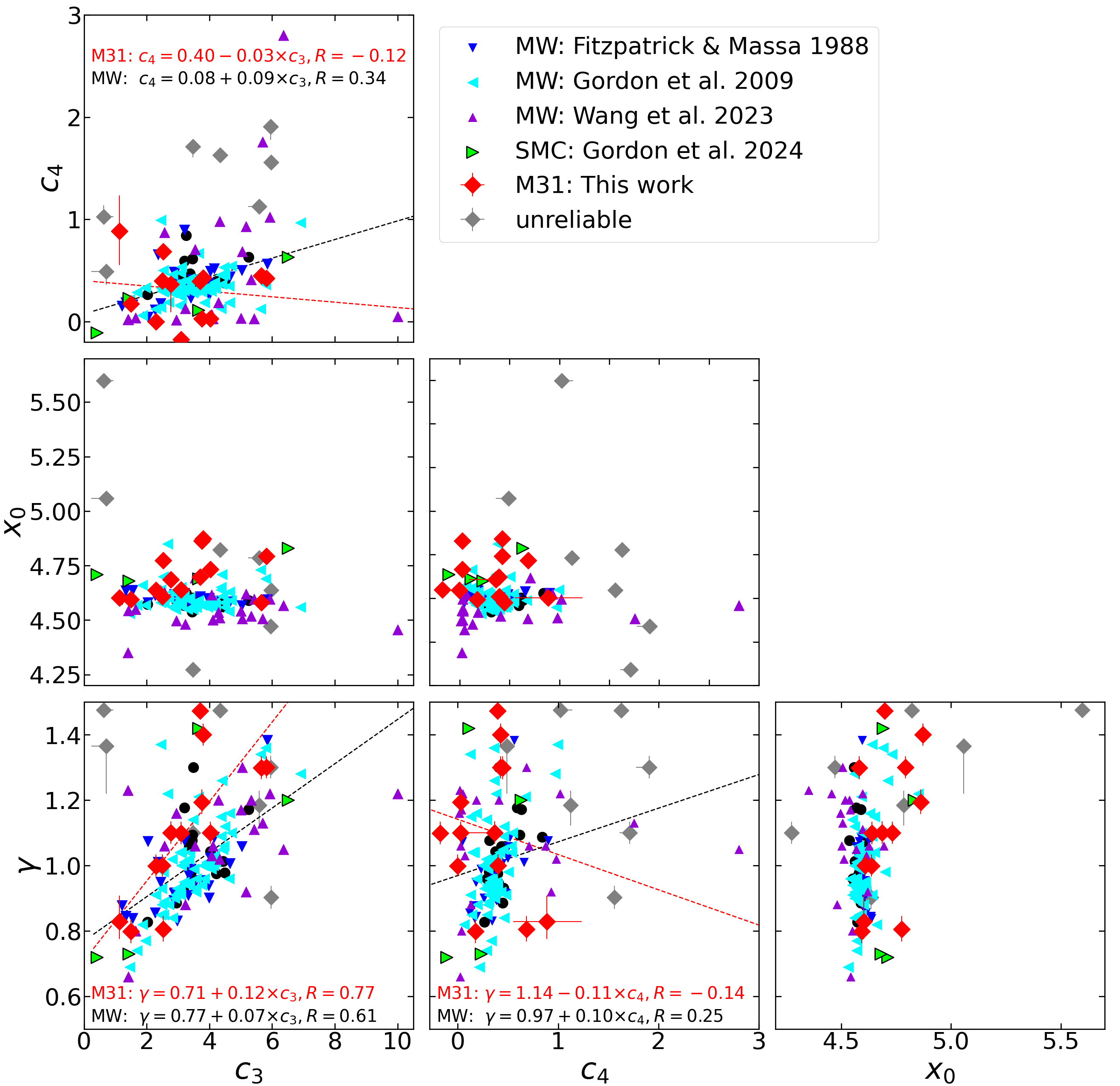}
    \caption{Relationships among $c_3$, $c_4$, $x_0$, and $\gamma$. Red diamonds show our M31 results, with gray diamonds marking low-S/N or poorly defined bumps. Blue, cyan, and purple triangles denote MW measurements from \citet{Fitzpatrick1988}, \citet{Gordon2009}, and \citet{Wang2023}, with black dots for overlapping sources; green triangles mark four SMC stars with a 2175 Å feature from \citet{Gordon2024}. In the $c_3$ VS $c_4$, $c_3$ VS $\gamma$, and $c_4$ VS $\gamma$ panels, red dashed lines indicate the best-fit relations for M31, black dashed lines the combined MW fits, and no SMC fit is shown due to the small sample size. The corresponding fitting coefficients and Pearson correlation values are given in the text. }
    \label{fig:params}
\end{figure}

\begin{figure}[htb!]
    \centering
    \includegraphics[width=0.8\textwidth]{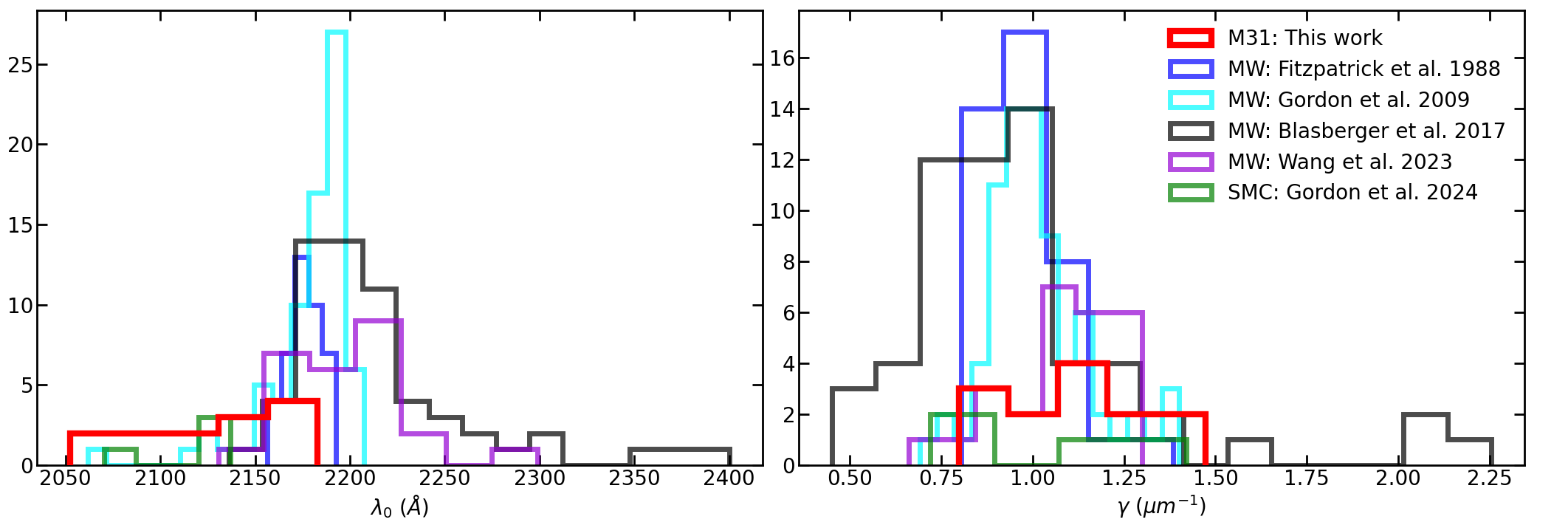}
    \caption{Comparison of the $\lambda_0$ and $\gamma$ parameter distributions derived in this work (red) with those reported by \citet{Fitzpatrick1988} (blue), \citet{Gordon2009}(cyan), \citet{Blasberger2017} (black), \citet{Wang2023} (purple), and \citet{Gordon2024}(green).}
    \label{fig:x0ga}
\end{figure}

\begin{figure}[htb!]
    \centering
    \includegraphics[width=0.7\textwidth]{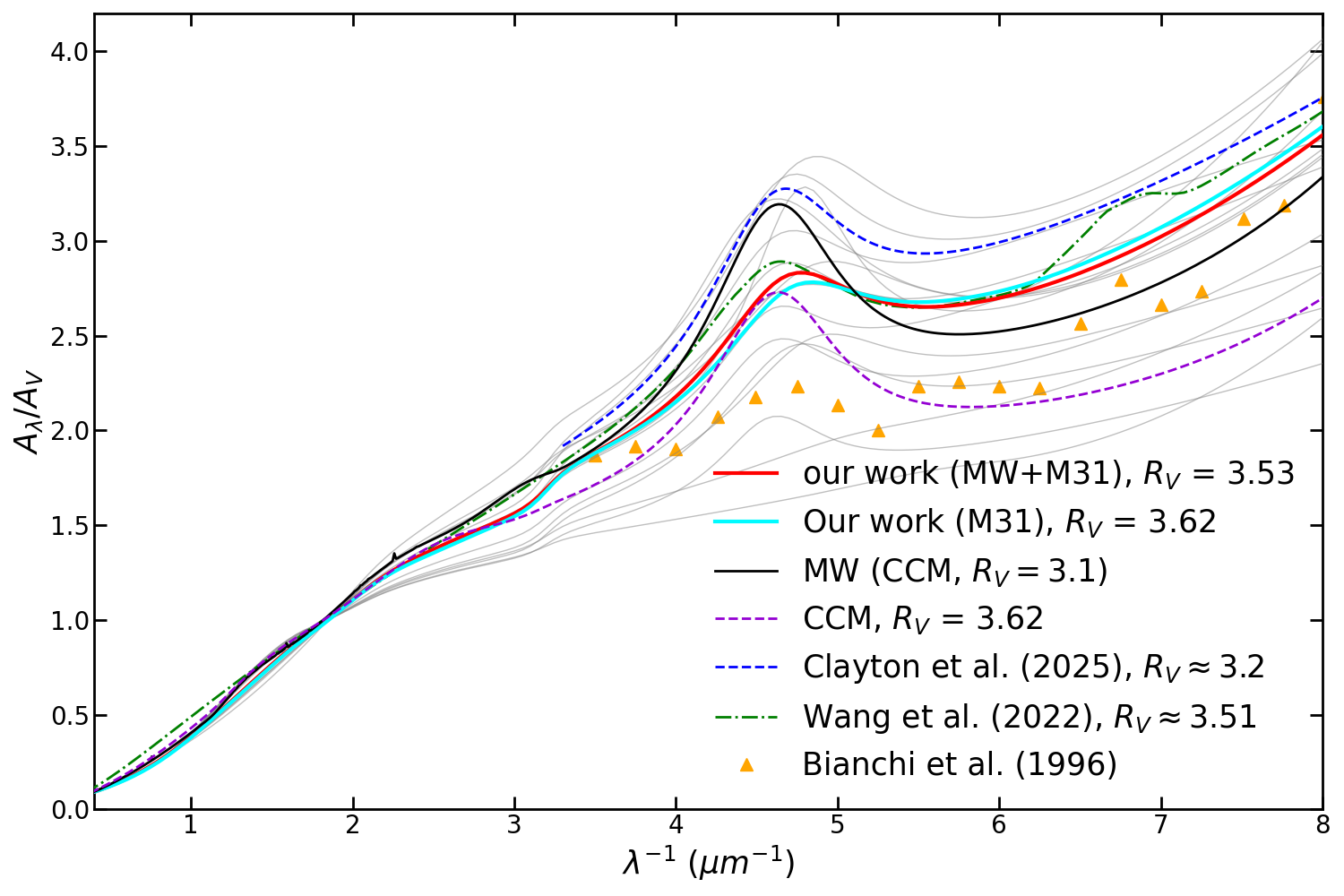}
    \caption{UV extinction curves for the 15 sightlines are shown as gray lines, with the red line denoting their average and the cyan line the M31 average after MW foreground correction. The purple dashed line shows the CCM curve with the same $R_V$ as our M31 average ($R_V=3.62$), the black line the average MW CCM curve, and the blue and green dashed lines the M31 averages from \citet{Clayton2025} and \citet{Wang2022}; orange triangles mark the flat curve from \citet{Bianchi1996}. Our average agrees with \citet{Wang2022} but departs from the CCM prediction in the UV, underscoring the model’s limitation.}
    \label{fig:average}
\end{figure}

\subsection{The Parameters of 2175 Å Feature}\label{sec:parameter}

Figure~\ref{fig:params} illustrates the relationships among the bump parameters ($c_3$, $c_4$, $x_0$, $\gamma$) derived in this work, compared with those reported by \citet{Fitzpatrick1988}, \citet{Gordon2009}, \citet{Wang2023} and \citet{Gordon2024}. For sources presented in multiple MW studies, we adopt averaged values for comparison. 
\citet{Fitzpatrick1988} found weak correlations among $c_3$, $c_4$, and $\gamma$ in the MW and suggested that both the 2175 Å bump and far-UV curvature may originate from resonant absorption by the same population of grains.
Consistent with earlier findings, our results show no correlation between $x_0$ and the other parameters. However, in contrast to \citet{Fitzpatrick1988}, we only detect a positive correlation between $c_3$ and $\gamma$, with a correlation coefficient $R =0.77$.

Figure~\ref{fig:x0ga} compares the distributions of $\lambda_0$ and $\gamma$ obtained here with those from \citet{Fitzpatrick1988}, \citet{Gordon2009}, \citet{Blasberger2017}, \citet{Wang2023} and \citet{Gordon2024}. While $\gamma$ displays a similar distribution across all studies, $\lambda_0$ in this work, \citet{Gordon2009} and \citet{Gordon2024} is systematically shifted toward shorter wavelengths, and in this work exhibits a broader spread. Seven sightlines have $\lambda_0$ between $2050-2150$ Å, and these deviations show no correlation with SNR. The fitting precision ($3-6$ Å) cannot account for the $\sim100$ Å offset. The excellent agreement between model and data supports the robustness of these results, implying that the allowable range for the 2175 Å central wavelength may extend beyond previously reported limits.

The diversity observed here parallels trends seen in other galaxies. For instance, the SMC was historically thought to lack the bump, yet \citet{Gordon2024} identified four SMC sightlines with MW-like curves, including a detectable 2175 Å feature. And the 2175 Å feature in SMC show excellent agreement with that in MW and M31. 
In this study, we find two sightlines toward M31 with negligible bump strength, reinforcing the complexity of interstellar extinction and the challenge of establishing a universal extinction law for correction purposes.

\subsection{The Average Extinction Curve toward M31}\label{sec:average curve}

Excluding the five low-SNR stars, the average extinction curve derived from the remaining 15 sightlines (red line), together with the individual extinction curves (gray lines), is presented in Fig.~\ref{fig:average}. Given the small sample size and the presence of two sightlines lacking a 2175 Å feature, we adopted the median to construct the average extinction curve. These sightlines exhibit a wide range of extinction behaviors, from flat profiles lacking the 2175 Å feature to steep curves with a pronounced bump. The average curve shows excellent agreement with that reported by \citet{Wang2022}. The UV portion of the average extinction curve ($R_V = 3.53$) can be described by the following analytical form:
\begin{equation}
    E(x-V)/E(B-V) = -1.03+1.1x+2.77D(x;1.0,4.697)+0.4F(x)
\end{equation}
or equivalently,
\begin{equation}
    A_x/A_V = 0.71+0.31x+0.78D(x;1.0,4.697)+0.11F(x)
\end{equation}
where $x\equiv\lambda^{-1}$. Details of the extinction model are provided in Section~\ref{sec:mod_ext}. 

Since our measurements trace the integrated extinction along each sightline, we compare the adopted MW foreground contribution based on \citet{Clayton2015} and \citet{Wang2022} with the MW extinction map toward M31 from \citet{Wang2025}. Adopting $A_V = E(B-V) \times R_V = 0.186$ mag for our sample yields an average absolute deviation of only $\sim 0.0035$ mag relative to the \citet{Wang2025} values, which is well below the $\sim 0.1$ mag precision of our fitting. Therefore, for consistency and simplicity, we adopt a fixed MW foreground extinction of $E(B-V) = 0.06$ mag \citep{Schlegel1998, Schlafly2011, Bianchi2012, ZhangRuoyi2020} with a CCM law of $R_V = 3.1$. We find that the MW foreground contributes approximately $A_\text{bump, MW} / A_\text{bump, total} = 33.2\%$ and $A_{V,\text{MW}} / A_{V,\text{total}} = 16.7\%$. The M31 average extinction curve after subtracting this foreground contribution ($R_V = 3.62$) is shown as the cyan solid line in Figure~\ref{fig:average}. Compared with a CCM curve of the same $R_V$, the corrected M31 curve agrees well in the optical–NIR regime but shows significant deviations in the UV, underscoring the limitations of the CCM prescription in reproducing UV extinction toward M31.

\section{Conclusions}
\label{sec:conclusion}
In this study, we investigated the extinction properties along sightlines toward M31 by analyzing UV spectra of bright O- and B-type supergiants obtained with HST. The main findings are summarized as follows:

1. The extinction curves exhibit substantial diversity, spanning from MW-like profiles to significantly flatter curves with higher $R_V$ values. The strength and prominence of the 2175 Å feature also vary widely, revealing that the dust properties along different sightlines in M31 are far from uniform. These variations likely reflect significant spatial inhomogeneities in dust composition and grain size distribution within the galaxy.

2. Among the bump parameters, $x_0$ shows no evident correlation with $c_3$, $c_4$, or $\gamma$, while a positive correlation is found between $c_3$ and $\gamma$. The central wavelength of the bump spans a broader range than previously reported, whereas the distribution of $\gamma$ remains consistent with earlier studies

3. The average extinction curve toward M31 corresponds to $R_V = 3.53$, consistent with previous studies. After the Galactic foreground contribution is removed, the average extinction curve yields $R_V = 3.62$. Nevertheless, its UV behavior shows clear deviations from the CCM model prediction, underscoring the limitations of a universal extinction law in characterizing extinction curves in M31. 

\normalem
\begin{acknowledgements}
This work is supported by the National Natural Science Foundation of china No. 12133002, 12373028, 12322304, 12173034, National Natural Science Foundation of Yunnan Province 202301AV070002 and Xingdian talent support program of Yunnan Province, and the Hebei NSF under grant No. A2023205036. This work has made use of data from the HST/STIS, LGGS, UKIRT, PS1 Survey, PHAT Survey, Swift/UVOT, and XMM-SUSS.

\end{acknowledgements}
  
\bibliographystyle{raa}
\bibliography{ms2025-0461}

\end{document}